\documentclass[%
 aip,
 amsmath,amssymb,
 reprint,%
]{revtex4-2}

\usepackage{graphicx}
\usepackage{dcolumn}
\usepackage{bm}
\usepackage{float}
\usepackage{color}
\usepackage[scanall]{psfrag} 

\usepackage[utf8]{inputenc}
\usepackage[T1]{fontenc}
\usepackage{mathptmx}
\usepackage{etoolbox}

\usepackage{amsmath, amsthm}
\usepackage{amsfonts}
\usepackage{soul}
\usepackage{enumerate}

\usepackage{caption}
\usepackage{subcaption}

\usepackage{CJKutf8} 

\usepackage{hyperref}
\hypersetup{
    colorlinks=true,
    linkcolor=blue,
    filecolor=magenta,      
    urlcolor=cyan,
    pdftitle={Overleaf Example},
    pdfpagemode=FullScreen,
    }

\newcommand{\Mod}[1]{\ (\mathrm{mod}\ #1)}

\newcommand{\dd}{\mathrm{d}}

\begin{document}
\title{Two-dimensional Brownian motion with dependent components: turning angle analysis}

\author{Micha{\l} Balcerek}%
    \email{Corresponding author: michal.balcerek@pwr.edu.pl}

    \affiliation{Faculty of Pure and Applied Mathematics, Hugo Steinhaus Center, Wroc{\l}aw University of Science and Technology, 50-370 Wrocław, Poland}%
    \affiliation{Department of Electrical and Computer Engineering and School of Biomedical Engineering, Colorado State University, Fort Collins, CO 80523, USA}

\author{Adrian Pacheco-Pozo}
\affiliation{Department of Electrical and Computer Engineering and School of Biomedical Engineering, Colorado State University, Fort Collins, CO 80523, USA}

\author{Agnieszka Wy{\l}omanska}
    \affiliation{Faculty of Pure and Applied Mathematics, Hugo Steinhaus Center, Wroc{\l}aw University of Science and Technology, 50-370 Wrocław, Poland}%

\author{Krzysztof Burnecki}
    \affiliation{Faculty of Pure and Applied Mathematics, Hugo Steinhaus Center, Wroc{\l}aw University of Science and Technology, 50-370 Wrocław, Poland}%

\author{Diego Krapf}
    \affiliation{Department of Electrical and Computer Engineering and School of Biomedical Engineering, Colorado State University, Fort Collins, CO 80523, USA}
\begin{abstract}
Brownian motion in one or more dimensions is extensively used as a stochastic process to model natural and engineering signals, as well as financial data.  Most works dealing with multidimensional Brownian motion {consider the different dimensions as independent components}. In this article, we investigate a model of correlated Brownian motion in $\mathbb{R}^2$, where the individual components are not necessarily independent. We explore various statistical properties of the process under consideration, going beyond the conventional analysis of the second moment. Our particular focus lies on investigating the distribution of turning angles. This distribution reveals particularly interesting characteristics for processes with dependent components that are relevant to applications in diverse physical systems. Theoretical considerations are supported by numerical simulations and analysis of two real-world datasets: the financial data of the Dow Jones Industrial Average and the Standard and Poor's 500, and trajectories of polystyrene beads in water. Finally, we show that the model can be readily extended to trajectories with correlations that change over time. 
\end{abstract}

\maketitle
\onecolumngrid 


Brownian motion is widely used to model systems that exhibit random fluctuations. The model is useful for describing the random movement of particles suspended in a fluid or any fluctuating signal. Often, these systems involve two or more dimensions, for example, when observing the behavior of small particles under a microscope or when evaluating multiple interconnected signals. In particular, the individual signals or components may depend on each other, such that this dependence alters the temporal evolution of the system.      This article describes processes in which the components of two-dimensional Brownian motion are correlated.  It is shown that a very useful statistical metric to analyze this process involves the distribution of turning angles, that is, the changes in direction between successive measured steps. The characterization using turning angles is employed to study both the motion of micrometer-sized spheres suspended in water and the fluctuations of financial indices. Furthermore, the method is also shown to be useful when the correlation between components is not constant but changes over time.  

\section{Introduction}
Since the pioneering works of Einstein and Smoluchowski \cite{einstein05,smoluchowski06}, Brownian motion (BM) has established itself as the quintessential stochastic model for diffusive transport \cite{mori1965transport,morters2010brownian,hanggi2005introduction}. BM is now widely applied in various disciplines, including biology \cite{saffman1975brownian,allen2010introduction}, statistical physics \cite{ebeling2005statistical}, chemistry \cite{hanggi1990reaction}, quantum mechanics \cite{schwinger1961brownian,friedman2017quantum}, ecology \cite{horne2007analyzing,vilk2022unravelling}, and finance \cite{meyer2003strategic}.
The multidimensional BM, also called the multidimensional Wiener process,  extends the classical one-dimensional BM to multiple dimensions, resulting in a vector of independent or correlated components. It is a fundamental stochastic process that is widely utilized to model the dynamics of systems that involve two or more variables. Each component of a multidimensional BM follows a normal distribution with stationary and independent increments, making it an appealing tool to capture the complexity of real-world phenomena. 
Multidimensional BM {is ubiquitous in science and technology. To name a few examples, multidimensional BM} was applied to model mobile radio channels \cite{6698114}, data in the aircraft industry \cite{6492037}, image encryption algorithms \cite{gao2021image,rawat2023new}, credit risk \cite{giesecke2004correlated,ching2021correlated}, diffusion of colloids \cite{hassan2015making} and protein diffusion in lipid membranes \cite{knight2009single,campagnola2015superdiffusive,krapf2016strange}. Typically, Brownian motion in $d$ dimensions is defined such that its projections along each dimension are statistically independent \cite{krapf2018power}. However, environmental constraints can introduce correlations between these different projections \cite{kou2016first,basu2018active}. {A typical example deals with financial time series, where two assets in a natural way are going in the same direction \cite{BIELAK2021102308}. However correlated Brownian motion appears in a very wide and diverse range of systems, a non-exhaustive list includes brain networks \cite{8831393}, two-dimensional motion of colloidal dimers \cite{mayer2021two}, optically trapped graphene flakes \cite{marago2010brownian}, Brownian motions on manifolds \cite{grong2024most}, and thermal motion of nanotubes \cite{tsyboulski2008translational,fakhri2010brownian}.}

Understanding the statistical properties of multidimensional BM is crucial to accurately represent and predict the behavior of multidimensional systems under the effect of random perturbations. In the literature, there are known works that explore the theoretical properties of multidimensional Brownian motion \cite{Follmer,MrongowiusOn2022,KendallCoupling2007,kuo1987generalized,Kou_Zhong_2016,TSEKOV1995175,SACERDOTE2016275}. The characteristics of various extensions of the multidimensional BM are also considered, such as multidimensional BM with membrane \cite{KOPYTKO2023371}, reflected multidimensional Brownian motion \cite{Blanchet_Murthy_2018}, multidimensional fractional BM \cite{sant2006text,Qian_1998,MarajZygmat,Ji2018ruin,MCGAUGHEY2002369}, multidimensional geometric and arithmetic BM \cite{Maruddani_2018,Domine_Pieper_1993}, multidimensional BM with jumps \cite{SACERDOTE201361}, and multidimensional Ornstein-Uhlenbeck process \cite{PhysRevE.99.062221,MarajZygmat,doi:10.1080/17442508.2024.2315274}. Despite the extensive range of applications and the popularity of multidimensional Brownian motion, the analysis of these processes when the individual components are correlated remains understudied. Thus, a more comprehensive characterization of these processes would enhance our understanding of the real phenomena under consideration.

This article focuses on a two-dimensional Brownian motion model where the motions in each dimension are correlated. This model is particularly relevant for systems where interactions or constraints induce such correlations, ranging from the movement of particles in a fluid to the dynamics of correlated financial assets. Understanding the behavior of Brownian motion with spatial correlations is crucial for the accurate modeling and prediction of the evolution of such systems.
The mathematical framework for our model is developed by defining the correlation between the two components through a correlation coefficient $\rho$. We derive essential properties, including the autocovariance and cross-covariance functions, as well as the mean square displacement. Going beyond these traditional statistical analyses, we study the turning angles generated by consecutive increments and discuss their significance.
Numerical simulations validate the theoretical findings and highlight the impact of the correlation between components on the distribution of the turning angles. Furthermore, we apply our model to empirical data from two vastly different fields: financial markets and physical systems, showcasing its versatility and practical applications.
We also introduce an extension of the discussed two-dimensional model that incorporates a time-varying correlation between the model's components, and present its theoretical formulation as well as practical applications. The effects of time-varying correlations are demonstrated using simulated data. Finally, the time-varying model is applied to historical financial data. We show that employing moving window methods analysis enables the detection and quantification of changes in the correlation coefficient over time, providing deeper insights into the underlying dynamics of the data.
Our study aims to enhance the understanding of multidimensional Brownian motion with spatial correlations, offering a robust framework for both theoretical analysis and practical applications. By comparing the theoretical properties of the model discussed with its statistical counterparts observed in the data, we provide a valuable tool for researchers and practitioners across various scientific disciplines. The insights gained from the analysis of the turning angles are particularly significant because of their independence from the frame of reference.

The remainder of the paper is organized as follows. We first present a generalized theoretical model of the correlated Brownian motion. We highlight the impact of the correlation coefficient $\rho$. Then, we emphasize the most important properties of the model, such as autocovariance and cross-covariance functions, mean square displacement, and we introduce the concept of turning angles. Theoretical calculations are verified using extended numerical simulations. Furthermore, we illustrate the applicability of the model on two datasets from various fields. Finally, we extend the model to incorporate the time-varying correlations and illustrate the usefulness of such an extension by fitting it to the real-world data. 

\section{Correlated BM Model}
Let us consider the following two-dimensional Brownian motion with spatial correlations
\begin{align}
    \mathbf{R}(t) = \begin{bmatrix}B_1(t)\\ 
    B_2(t)\end{bmatrix},
\end{align}
where $t\geq 0$ and $B_1(t)$ and $B_2(t)$ are spatially correlated one-dimensional Brownian motions with spatial correlation coefficient $\rho \in (-1, 1)$, and diffusion coefficients $D_1$ and $D_2$, respectively. One could view the two Brownian motion components as
\begin{align}
    \label{eq:bm_construction}
    \begin{bmatrix}
        B_1(t)\\
        B_2(t) 
    \end{bmatrix} = 
    A \begin{bmatrix}
        W_1(t)\\
        W_2(t) 
    \end{bmatrix},
\end{align}
where $W_1(t)$ and $W_2(t)$ are independent one-dimensional Brownian motions, and $A = \begin{bmatrix}
    \sqrt{2D_1} & 0\\
    \sqrt{2D_2}\rho & \sqrt{2D_2(1-\rho^2)}
\end{bmatrix}$. The parameter $\rho$ controls the strength of the linear dependence between coordinates and their increments. Matrix $A$, sometimes referred to as the Cholesky triangle \cite{golub2013matrix} 
obtained from the Cholesky decomposition of the matrix $\Sigma$, is chosen in such a way that $\Sigma = A A^T$ is the covariance matrix of $\mathbf{R}(1)$, i.e.
\begin{align}
    \Sigma = \begin{bmatrix}
        2D_1 & 2\rho \sqrt{D_1 D_2}\\
        2\rho \sqrt{D_1 D_2} & 2D_2
    \end{bmatrix}.
    \label{eq:cov_matrix}
\end{align}

\begin{figure}[ht!]
\centering
\includegraphics[width=0.9\linewidth]{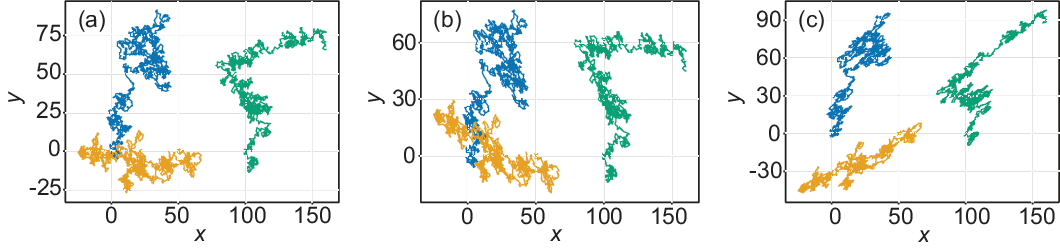}
\caption{Representative numerical simulations of two-dimensional Brownian motion trajectories with different correlations. In all panels $D_1 = D_2 = 0.5$, time step $1$ and trajectory length $2^{11}$. (a) Trajectories with $\rho=0$, (b) $\rho=-0.3$ and (c) $\rho=0.6$. Each trajectory is shown with a different color (yellow, blue, or green). For clarity of presentation, trajectories were shifted by 50 (yellow) and 100 units (green) to the right. The underlying randomness is the same for all panels {with paths with the same color corresponding to equivalent trajectories}.}
 \label{fig:trajs1}
\end{figure}

Figure \ref{fig:trajs1} presents sample trajectories of this model, where we use the notation $x(t)=B_1(t)$ and $y(t)=B_2(t)$ as this intuitively exemplifies motion in a plane. The independent case ($\rho=0$) is presented in Figure \ref{fig:trajs1}(a), a negative dependent case with $\rho=-0.3$ is shown in Figure \ref{fig:trajs1}(b), and a positive dependent case with $\rho=0.6$ in Figure \ref{fig:trajs1}(c). In the three presented cases, we used $D_1=D_2 = 0.5$, time step $1$, and trajectory length of $2^{11}$ data points. To highlight the influence of the correlation coefficient $\rho$, we used the same random noise in all the panels, that is, the underlying randomness corresponding to the trajectory in every color is the same.

\begin{figure}[ht!]
\centering
\includegraphics[width=0.9\linewidth]{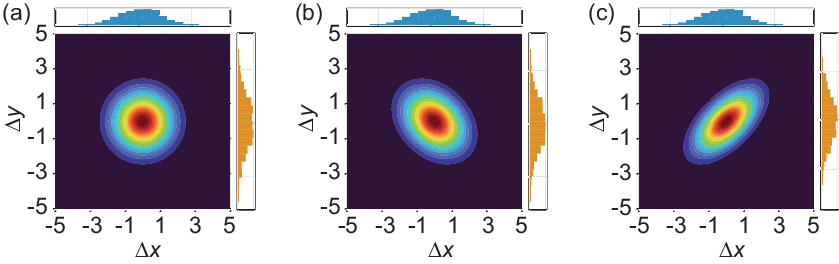}
\caption{Contour plots of two-dimensional joint PDFs of the increments of BM with spatial correlations. The histograms on the top and right of the joint PDFs correspond to the increments of each individual component. (a) $\rho=0$, (b) to $\rho=-0.3$ and (c) to $\rho=0.6$.}
 \label{fig:densities2}
\end{figure}

\section{Properties of the correlated BM with spatial correlations}
In this section, we present the basic properties of the two-dimensional Brownian motion with spatial correlations. {Physical measurements are naturally sampled at discrete times, thus henceforth in the article we consider the process to be discretized at times $t=0,1,2,...$.}
First, let us analyze the dependence structure of increments $\Delta^\delta \mathbf{R}(t) \equiv \mathbf{R}(t+\delta) - \mathbf{R}(t)$. We will also use the notation $\Delta^1 \equiv \Delta$ for increments {corresponding to $\delta=1$}. Let us first consider the dependence structure on the same coordinate. The autocovariance function of such increments, closely related to the velocity autocorrelation function (VACF)\cite{qian1991single}, is given by \cite{karatzas2014brownian}
\begin{align}
    \gamma_{\Delta^\delta B_j, \Delta^\delta B_j}(h) \equiv \langle \Delta^\delta B_j(t) \Delta^\delta B_j(t+h) \rangle = \begin{cases}
        2D_j (\delta - h) \quad &\textrm{ for } h < \delta,\\
        0 \quad &\textrm{ for } h \geq \delta,
    \end{cases}
    \label{eq:vacf1}
\end{align}
for $j = 1, 2, h\geq0$, and $\Delta^\delta B_j(t) = B_j(t+\delta)-B_j(t)$. The expressions for $h< \delta$ and $h \geq \delta$ correspond to overlapping and disjoint increments, respectively. For example, considering $\delta=1$ and $h\in \mathbb{Z}_+$ for the discrete times $t=0, 1, \ldots$, we have 
\begin{align}
    \gamma_{\Delta B_j, \Delta B_j}(h) \equiv \langle \Delta B_j(t) \Delta B_j(t+h) \rangle = \begin{cases}
        2D_j \quad &\textrm{ for } h = 0,\\
        0 \quad &\textrm{ for } h \neq 0.
    \end{cases}
\end{align}
Since we deal with a two-dimensional model, it is not sufficient to consider only the covariance structure on one coordinate. Taking both coordinates into consideration leads to the so-called cross-covariance
\begin{align}
    \gamma_{\Delta^\delta B_1, \Delta^\delta B_2}(h) \equiv \langle \Delta^\delta B_1(t) \Delta^\delta B_2(t+h) \rangle = \begin{cases}
        2\rho \sqrt{D_1 D_2} (\delta - h)\quad &\textrm{ for } h < \delta,\\
        0 \quad &\textrm{ for } h \geq \delta.\\
    \end{cases}
    \label{eq:cross_cov}
\end{align}
{Note that eqs.~(\ref{eq:vacf1}) and (\ref{eq:cross_cov}) are also valid for continuous time.} One can easily show that such a function is symmetric in the sense $\gamma_{\Delta^\delta B_1, \Delta^\delta B_2}(h) = \gamma_{\Delta^\delta B_2 \Delta^\delta B_1}(h)$.
Due to the construction given in eq.~(\ref{eq:bm_construction}), the process is Gaussian and the increments $\Delta^\delta \mathbf{R}(t)$ have a Gaussian distribution with zero mean and covariance structure given by \begin{align}
\label{eq:sigam_time}
\Sigma(\delta) = \begin{bmatrix}
    2D_1 \delta & 2\sqrt{D_1 D_2}\rho \delta\\
    2\sqrt{D_1 D_2}\rho \delta & 2D_2 \delta\\
\end{bmatrix}.
\end{align}

Figure \ref{fig:densities2} shows the influence of the correlation $\rho$ on the joint probability density function (PDF) of the BM increments. Note that all histograms on the top and right sides of Figures \ref{fig:densities2}(a), \ref{fig:densities2}(b), and \ref{fig:densities2}(c) depict the same Gaussian distribution, not influenced by $\rho$. However, the joint distributions presented on the contour plots are significantly different. Figure \ref{fig:densities2}(a) presents the independent case ($\rho=0$), while the others show the correlated cases: Figures \ref{fig:densities2}(b) the low negative one ($\rho=-0.3$), and Figure \ref{fig:densities2}(c) the high positive one ($\rho=0.6$). 
Note that the correlation coefficient $\rho$ is highly dependent on the reference frame. For example, in the cases of the positive and negative dependence, mirroring the $Ox$ axis (i.e. reflection about the vertical axis) would result in the opposite sign of their corresponding correlation coefficients $\rho$. Even in the independent case $\rho=0$, a non-zero correlation can emerge when the frame of reference is rotated by any non-right angle.

Another property that is usually analyzed is the ensemble mean square displacement (MSD). Taking both coordinates into consideration, we get
\begin{align}
    \langle R^2(t) \rangle \equiv \langle B_1^2(t) + B_2^2(t) \rangle = 2D_1 t + 2D_2 t = 2(D_1 + D_2) t,
\end{align}
where $R(t) = |\mathbf{R}(t)|$ is the magnitude of $\mathbf{R}(t)$ and $R^2(t) = |\mathbf{R}(t)|^2$ is its square.

\begin{figure}[ht!]
\centering
\includegraphics[width=0.5\textwidth]{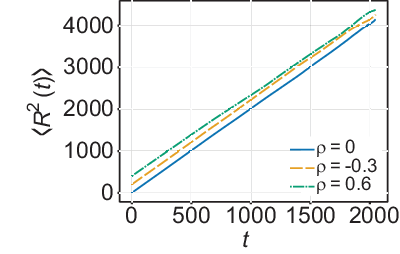}
\caption{Ensemble mean square displacement of correlated BM. The solid blue line corresponds to the case with independent components ($\rho=0$), dashed orange to the negatively correlated case ($\rho=-0.3$), and the dash-dotted green line to the positively correlated case ($\rho=0.6$). {For clarity,} the MSDs corresponding to $\rho=-0.3$ and $\rho=0.6$ are shifted by 200 and 400, respectively.}
 \label{fig:msd3}
\end{figure}

It should be noted that, despite introducing the spatial correlation, $\rho$ does not influence MSD. The model still describes normal diffusion, just like each of its components.
Figure \ref{fig:msd3} presents the MSD based on simulated 1000 trajectories of two-dimensional BM. The three considered cases ($\rho=0, \rho = -0.3$, and $\rho=0.6$ corresponding to, in order, the solid blue line, the dashed orange line, and the dashed-dotted green line) exhibit linear behavior. The lines describing the dependent cases are shifted by a constant of 200 (dashed orange line) and 400 (dashed-dotted green line) for visual clarity.

\subsection{Polar representation} 
\label{sseq:angles}
In this subsection, we present the obtained results related to the angles $\alpha_t$ generated by the process $\mathbf{R}(t)$.
From the properties presented previously, we know that $\mathbf{R}(t)$ has a multivariate normal distribution with zero mean and covariance matrix $\Sigma(t)$. 
Utilizing the polar representation $(R(t), \alpha_t)$ of the position $\mathbf{R}(t)$ we can obtain the polar angles $\alpha_t$ by the following relation 
\begin{align}
\mathbf{R}(t)/R(t) = \begin{bmatrix}B_1(t)/R(t)\\ B_2(t)/R(t)\end{bmatrix} = \begin{bmatrix}\cos \alpha_t\\ \sin \alpha_t\end{bmatrix}.
\end{align} 
Thus, the PDF of the angles $\alpha_t$ generated with the horizontal axis is \cite{wang2014modeling, hernandez2017general} 

\begin{align}
\label{eq:angles_general}
    p_{\alpha}(\theta; t) = \frac{1}{2\pi \sqrt{|\Sigma(t)|} \begin{bmatrix} \cos\theta \\
    \sin\theta\end{bmatrix}' \cdot \Sigma(t)^{-1} \cdot \begin{bmatrix} \cos\theta \\
    \sin\theta\end{bmatrix}}, \quad \theta \in (0, 2\pi).
\end{align}
%
{We note that $\theta$ is simply used as the argument for the PDF of the angles $\alpha$, and that $\alpha$ is defined in the interval $(0,2\pi)$ because they refer to the polar angle of the position of the tracer (see Figure~\ref{fig:turning_angles4}).} Using the explicit form of matrix $\Sigma(t)$ given in eq.~(\ref{eq:sigam_time}), we can write
\begin{align}
    \label{eq:angle_explicit}
    p_\alpha(\theta) = \frac{\sqrt{1-\rho^2}}{2\pi} \frac{1}{\cos^2(\theta) \sqrt{\frac{D_2}{D_1}} + \sin^2(\theta) \sqrt{\frac{D_1}{D_2}} - 2\rho \sin(\theta)\cos(\theta)}, \qquad \theta \in (0, 2\pi),
\end{align}
where we omitted the time variable $t$ in $p_\alpha(\theta)$, as the PDF of the angles on the right-hand side of the equation does not depend on it. 

Now, assuming that we have the same diffusivity on both coordinates, i.e. $D_1 = D_2 = D$, eq. (\ref{eq:angle_explicit}) further simplifies to
\begin{align}
    p_\alpha\left(\theta\right) = \frac{\sqrt{1-\rho^2}}{2\pi}\frac{1}{1-\rho \sin(2\theta)}, \quad \theta \in (0, 2\pi).
\end{align}
We highlight that by construction, the distribution of angles $\alpha_t$ depends on the chosen frame of reference, which, in our case, is the horizontal axis. In many applications, such a choice is arbitrary and, thus, might be undesirable. Thus, in order to avoid dependence on the choice of reference system, we employ the information provided by consecutive observations.

To do so, we first introduce the angles $\alpha_t^\Delta$ generated by increments $\Delta^\delta \mathbf{R}(t)$. To simplify the notation, we shall henceforth consider a unitary time step $\delta=1$. The presented distribution of angles does not change, because only the covariance matrix differs in time-scaling, which does not influence the resulting distribution in any way. Additionally, in the case of our model, due to the stationarity of the increments, and more importantly their independence (provided that the increments are taken over disjoint intervals), we have independence between angles $\alpha_t^\Delta$ for different $t$. 


\subsection{Turning angles}
\label{ssec:turning_angles}
\begin{figure}[ht!]
\centering
\includegraphics[width=0.6\linewidth]{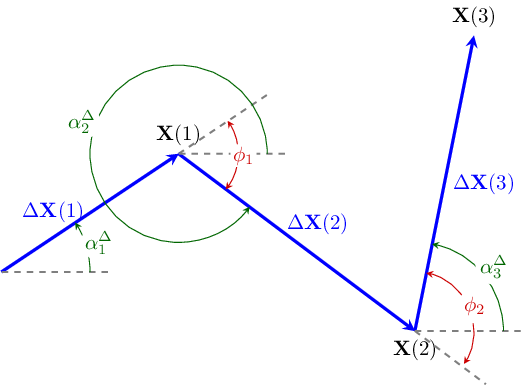}
\caption{{Schematic representation of the polar angles $\alpha_t^\Delta$ (marked with dark green lines) and turning angles $\phi_t$ (marked with dark red lines).}}
\label{fig:turning_angles4}
\end{figure}

While the information given by the distribution of angles $\alpha_t$ is interesting, it is not necessarily useful in the case of real data, where the choice of any specific axis might be arbitrary. 
A more natural approach is to consider the so-called turning angles, generated by consecutive increments \cite{codling2008random,burov2013distribution,bos2015angular,sadegh2017plasma,kadoch2017directional,mosqueira2020antibody,fang2022time,hidalgo2024directed}. In the correlated BM, the increments are still independent. This independence, as stated in the previous section, leads to the independence of the corresponding angles $\alpha_t^\Delta$. 
Specifically, let $\phi_t$ denote the angle between two consecutive increments at times $t$ and $t+1$. Figure \ref{fig:turning_angles4} shows the angles of the increments $\alpha_t^\Delta$ and the turning angles $\phi_t$ in a sketched interval consisting of three consecutive observations within a trajectory. Angles $\alpha_t^\Delta$ are measured to the horizontal axis, preserving the counter-clockwise direction (cf. $\alpha_2^\Delta$ in the figure). The turning angles $\phi_t$ are defined by
\begin{align}
    \label{eq:angle_def}
    \cos(\phi_t) = \frac{\Delta \mathbf{R}(t+1) \cdot \Delta \mathbf{R}(t)}{|\Delta \mathbf{R}(t+1)| |\Delta \mathbf{R}(t)|},
\end{align}
{This definition of turning angles is directly connected to the velocity autocorrelation function, albeit without the ensemble average \cite{burov2013distribution}. While in two dimensions it is possible to work with $\phi$ on the interval $[0, 2\pi)$, the definition on the interval $[0, \pi)$, which arises from inverting eq.~(\ref{eq:angle_def}), can be directly extended to higher dimensions \cite{fang2022time}. If a process involves torque or chirality, then the full circle should be considered but that is not the case here. For these reasons, most of the literature on turning angles currently adopts the interval  $[0, \pi)$ \cite{sadegh2017plasma,kadoch2017directional,mosqueira2020antibody,fang2022time,hidalgo2024directed} and we adhere to this convention.} The connection to the angles $\alpha_t^\Delta$, described in the previous section, is as follows
{
\begin{align}
    \label{eq:angle_diff}
  \phi_t \equiv \begin{cases}
        |\alpha_{t+1}^\Delta-\alpha_t^\Delta|, \quad & \mathrm{for}\  \alpha_{t+1}^\Delta-\alpha_t^\Delta \in (-\pi,\pi),\\
        \\
        2\pi-(\alpha_{t+1}^\Delta-\alpha_t^\Delta), \quad \quad &\mathrm{for}\  \alpha_{t+1}^\Delta-\alpha_t^\Delta \in (\pi,2\pi),\\
        \\
        2\pi+(\alpha_{t+1}^\Delta-\alpha_t^\Delta), \quad \quad &\mathrm{for}\  \alpha_{t+1}^\Delta-\alpha_t^\Delta \in (-2\pi,-\pi).
    \end{cases}
\end{align}
From the properties of Brownian motion, the distribution of turning angles is symmetric about $\pi$, that is $g_\phi(\theta)=g_\phi(\pi-\theta)$ (this will also be verified later). Thus, we can employ a modified variable $\hat\phi$ that has the same distribution as $\phi$,
    \begin{align}
        \label{eq:angle_mod}
        \hat\phi_t \equiv \alpha_{t+1}^\Delta-\alpha_t^\Delta \Mod{\pi}.
    \end{align}
which makes it easier to analytically compute the distributions. In the rest of the article, we use $\hat\phi$ in all analytical derivations and $\phi$ in all numerical simulations. Both distributions are indeed found to be the same.}

For correlated BM, we can calculate the PDF of $\phi_t$ directly. Due to the increments of Brownian motion being independent and identically distributed (i.i.d.), we know that $\alpha_t^\Delta$'s are independent, and both $\alpha_t^\Delta$'s and $\phi_t$'s are identically distributed, and their distribution does not depend on time. This is not universally true and would not be the case for, for example, a two-dimensional fractional Brownian motion. Considering correlated BM, we can calculate the convolution of the PDF given in eq. (\ref{eq:angles_general}).  
%
Then the density of $\alpha_{t+1}^\Delta - \alpha_t^\Delta$, prior to the modulo operation, is given by
{ 
\begin{align}
    \label{eq:convolution}
    g_{{\hat\phi}}(\theta) &= (p_\alpha * \overline{p}_{-\alpha})(\theta) \nonumber \\
    &= \frac{1-\rho^2}{4\pi^2} \int_{\max\{0, \theta\}}^{\min\{2\pi, 2\pi+\theta\}} \frac{1}{\cos^2({x}) \sqrt{\frac{D_2}{D_1}} + \sin^2({x}) \sqrt{\frac{D_1}{D_2}} - 2\rho \sin({x})\cos({x})}
    \times \nonumber\\
    &\hphantom{=\frac{1-\rho^2}{4\pi^2} \int_{\max\{0, \theta\}}^{\min\{2\pi, 2\pi+\theta\}}} \frac{1}{\cos^2(\theta-x) \sqrt{\frac{D_2}{D_1}} + \sin^2(\theta-x) \sqrt{\frac{D_1}{D_2}} - 2\rho \sin(\theta-x)\cos(\theta-x)}\mathrm{d}x, \quad {\theta \in(-2\pi, 2\pi).}
\end{align}
}
 
For the special case of the same diffusivity for both coordinates ($D_1=D_2$), we have
\begin{align}
    \label{eq:convolution_special}
    g_{{\hat\phi}}(\theta) &= (p_{\alpha} * \overline{p}_{-\alpha})(\theta)  
   = \frac{1-\rho^2}{4\pi^2} \int_{\max\{0, \theta\}}^{\min\{2\pi, 2\pi+\theta\}} \frac{1}{1-\rho \sin(2x)} \frac{1}{1+\rho \sin(2(\theta-x))}\mathrm{d}x,
\end{align}
where $\overline{p}_{-\alpha}(\theta) = p_\alpha(-\theta), \mathrm{ for}\ \theta \in (-2\pi,0)$, is the density of $-\alpha$, i.e.
\begin{align*}
    \overline{p}_{-\alpha}(\theta) = \frac{\sqrt{1-\rho^2}}{2\pi} \frac{1}{1+\rho \sin(2\theta)}, \quad \theta \in (-2\pi, 0).
\end{align*}
{If we again relax the assumption that $D_1=D_2$, for any $D_1, D_2>0$, we take} into consideration the modulo operation, we arrive at the final formula for the density of $\phi_t$
\begin{align}
\label{eq:angle_mod_final}
    p_{{\hat\phi}}(\theta) = g_{{\hat\phi}}(\theta) +  g_{{\hat\phi}}(\theta + \pi) + g_{{\hat\phi}}(\theta - \pi) + g_{{\hat\phi}}(\theta - 2\pi).
\end{align}

Let us consider one more specific case, namely, uncorrelated coordinates with the same {diffusion coefficient}, i.e. the isotropic case.
In such a case, we have
\begin{align*}
    p_\alpha(\theta) &= \frac{1}{2\pi}, \quad \theta \in (0, 2\pi),\\
    \overline{p}_{-\alpha}(\theta) &= \frac{1}{2\pi}, \quad \theta \in (-2\pi, 0),
\end{align*}
and, via simple calculations
\begin{align}
    \label{eq:g_isotropic}
    g_{{\hat\phi}}(\theta) = \begin{cases}
        \frac{1}{(2\pi)^2} \theta + \frac{1}{2\pi}, \quad \theta \in (-2\pi, 0),\\
        \frac{-1}{(2\pi)^2} \theta + \frac{1}{2\pi}, \quad \theta \in (0, 2\pi).
    \end{cases}
\end{align}
Having the formula for $g_{{\hat\phi}}$ in eq. (\ref{eq:g_isotropic}), we can plug it into eq. (\ref{eq:angle_mod_final}) to obtain the PDF after modulo operation. Indeed, for $\theta \in (0,\pi)$ we have
\begin{alignat*}{4}
    p_{{\hat\phi}}(\theta) 
    &= g_{{\hat\phi}}(\theta) &&+  g_{{\hat\phi}}(\theta + \pi) &&+ g_{{\hat\phi}}(\theta - \pi) &&+ g_{{\hat\phi}}(\theta - 2\pi) \\ 
     &= \frac{-1}{(2\pi)^2}\theta &&+ \frac{-1}{(2\pi)^2}(\theta +\pi) &&+ \frac{1}{(2\pi)^2}(\theta - \pi) &&+ \frac{1}{(2\pi)^2}(\theta - 2\pi) + 4\times \frac{1}{2\pi} \\
     & = \frac{1}{\pi}.
\end{alignat*}
As expected, this is the PDF of a uniform distribution in the interval $(0, \pi)$. This is not the case when the conditions $\rho=0$ or $D_1=D_2$ are not met.

\begin{figure}[ht!]
\centering
\includegraphics[width=0.6\linewidth]{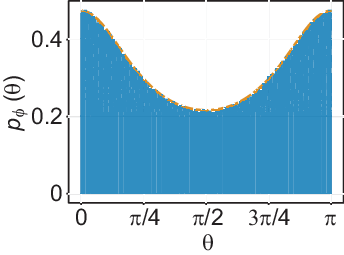}
\caption{Distribution of turning angles $\phi_t$ for the process with $D_1 = 2$, $D_2 = 1$, and $\rho = 0.7$. The solid blue histogram corresponds to the angles calculated using eq. (\ref{eq:angle_def}) from simulated data, and the dashed yellow line corresponds to {$p_{\hat\phi}$ from}~eq.~(\ref{eq:angle_mod_final}).}
\label{fig:turning_angles_density5}
\end{figure}

For comparison, we consider  $D_1 = 2, D_2 = 1$ and $\rho=0.7$. The PDF of the turning angles is presented in Figure~\ref{fig:turning_angles_density5}. We can see a clear deviation from the uniform distribution. {The solid blue histogram represents the turning angles derived from the simulated data} (based on 1000 trajectories of length $2^{10}$ with {$\delta=1$}), while the dashed yellow line corresponds to eq.~(\ref{eq:angle_mod_final}).

\section{Data}
\begin{figure}[ht!]
\centering
\includegraphics[width=0.9\linewidth]{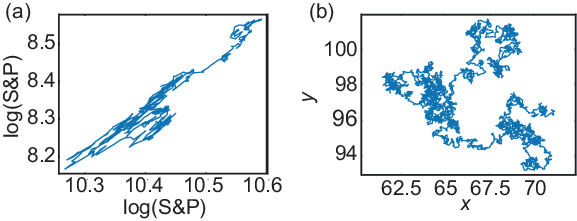}
 \caption{{Plots of representative sample trajectories of the analyzed datasets. (a) Financial data. (b) Beads in water.}}
     \label{fig:data_trajs}
\end{figure}

In this section, we present the results for two vastly different datasets. The first is a financial dataset that encompasses two of the main US market indices, the Dow Jones Industrial Average (DJIA) and the Standard and Poor's 500 (S\&P 500), for which we analyze their logarithmic returns. The second one relates to a soft-matter physical system, namely the motion of polystyrene beads in water. {In Figure \ref{fig:data_trajs}(a) we present the two-dimensional time series of the financial data considered, while in Figure \ref{fig:data_trajs}(b) we demonstrate the example two-dimensional trajectory of polystyrene beads in water.}
Despite obtaining universal results for any {diffusion coefficients} $D_1, D_2$ related to different directions, we first choose to normalize each of the coordinates considered by their corresponding standard deviation. {While this normalization is not useful in physical systems such as the motion of a particle, it is natural in cases where a consistent unit system is not well defined, such as financial indices. In this case, normalization makes the system of units consistent. For a physical system, the normalization procedure is superfluous and is equivalent to non-dimensionalization.}

\subsection{Example: Financial data}

The S\&P 500 is widely recognized as the main benchmark for large-cap U.S. equities. It comprises 500 leading companies from all sectors of the U.S. stock market, representing approximately 80\% of the U.S. equity market capitalization and over 50\% of the global equity market.
The DJIA is the second-oldest among U.S. market indices and most well-known stock index in the U.S. It was initially created to track 12 of the nation's largest corporations, and the index today consists of 30 prominent blue-chip stocks.
Both the DJIA and the S\&P 500 focus on large-cap U.S. stocks. The DJIA includes well-established and major companies often referred to as blue chips. Similarly, the S\&P 500 features top companies from key industries within the large-cap market segment. All DJIA stocks are typically included in the S\&P 500, where they generally make up between 25\% and 30\% of their market value.
Historically, the performance of the DJIA and the S\&P 500 have been closely correlated, often rising and falling in response to the same market forces. This correlation is expected due to their similar market exposures and comparable, though not identical, levels of volatility. However, there are notable differences in their performance, reflecting variations in their composition and investment styles.

\begin{figure}[ht!]
\centering
\includegraphics[width=0.9\linewidth]{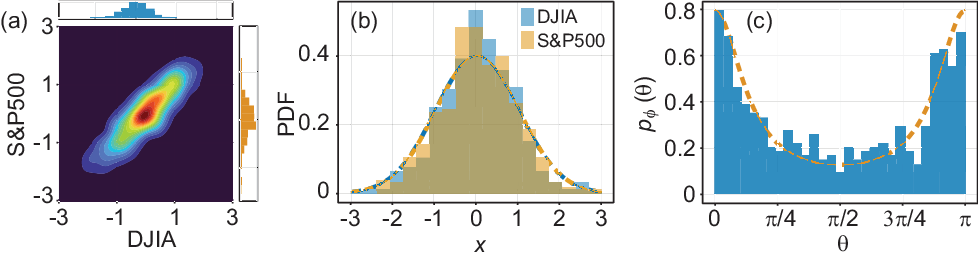}
\caption{Data analysis of the logarithmic returns of DJIA and S\&P500 indices measured in the span of 2 years, from May 1, 2022, to May 1, 2024. (a) 
     Contour plot of the two-dimensional density of the logarithmic returns, together with their histograms (DJIA on top, S\&P500 on the right side of the plot). (b) Histograms of the logarithmic returns with the fitted normal distributions. Blue bars and dashed blue line correspond to DJIA and yellow bars and dashed yellow line correspond to S\&P500. (c) Empirical distribution of turning angles. The dashed yellow line corresponds to the theoretical density (eqs.~(\ref{eq:convolution}) and (\ref{eq:angle_mod_final})).}
     \label{fig:data_finance}
\end{figure}

In Figure \ref{fig:data_finance} we present the analysis of the logarithmic returns of these two indices measured daily for a period of 2 years, from May 1, 2022, to May 1, 2024. Figure \ref{fig:data_finance}(a) shows the joint PDF, with the top and right histograms corresponding to DJIA and S\&P500 logarithmic returns, respectively. In Figure \ref{fig:data_finance}(b), we present fitted Gaussian PDFs of the logarithmic returns of both indices over the selected period. The blue bars and dashed blue lines correspond to the DJIA index, while the yellow bars and dashed-dotted yellow lines correspond to S\&P500. In Figure \ref{fig:data_finance}(c) we compare the empirical PDF of turning angles (blue bars) to the analytical density given by eq. (\ref{eq:angle_explicit}). 
{These results show a strong correlation between the two evaluated components, the DJIA and S\&P500 logarithmic returns, as expected for two indices both focusing on U.S. stocks. These correlations serve to validate the presented approach where the turning angle distribution agrees well with our analytical predictions.}

\subsection{Example: Polystyrene beads in water}
\begin{figure}[ht!]
\centering
\includegraphics[width=0.9\linewidth]{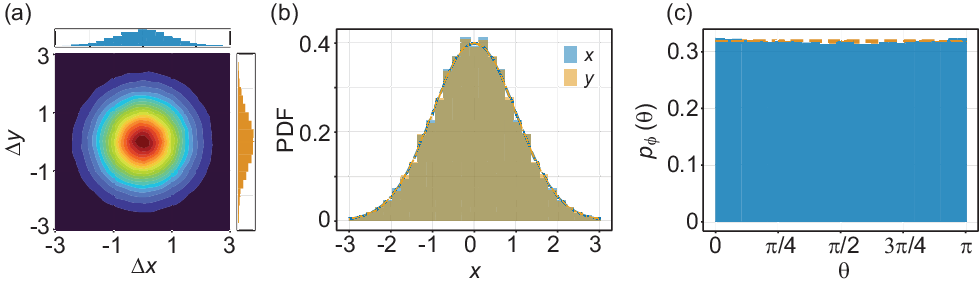}
\caption{Data analysis of polystyrene beads in water. (a) Contour plot of the two-dimensional joint PDF of the bead displacements, together with histograms ($x$ displacement on top, $y$ displacement on the right) for one single trajectory. (b) Histograms of all the displacements with the fitted normal distributions. Blue bars and dashed blue line correspond to $x$ displacements and yellow bars and dashed yellow line correspond to $y$ displacements. (c) Empirical distribution of turning angles. The dot-dashed yellow line corresponds to the theoretical density (eqs~(\ref{eq:convolution}) and (\ref{eq:angle_mod_final})).}
     \label{fig:7}
\end{figure}

The data consists of 150 two-dimensional trajectories of polystyrene beads submersed in water imaged during $40.96$ s at a frame rate of $100$ frames/s \cite{krapf2018power,thapa2021leveraging}. The nominal bead radius is $0.6\ \mu m$.
The movement of the beads can be modeled using a classical Brownian motion without correlations between the coordinates ($\rho=0$). Due to the isotropic nature of the environment, we observe that $D_1$ and $D_2$ are indeed the same. The density of increments (presented in Figure \ref{fig:7}(a)) has the circular contour lines of the characteristic of the joint PDF of two i.i.d. random variables. The one-dimensional histograms of the data are presented in Figure \ref{fig:7}(b) and compared with the Gaussian PDFs with matched estimated parameters ($\langle D\rangle= 0.375\mu\textrm{m}^2$/s). In this case, given $\rho=0$, the turning angles appear to have a uniform distribution (Figure \ref{fig:7}(c)). {In this case the characterization of the projections along the $x$ and $y$ directions, in terms of the turning angle distribution unequivocally shows the two components are not correlated, in agreement with the analytical predictions. The visual inspection of the 2D histogram also validates these conclusions.}

\section{Time-varying correlation}
{In many real-world data, additional complexities can cause the diffusion process to change over time due to temporal or spatial heterogeneities \cite{montiel2006quantitative,akimoto2017detection,pacheco2024fractional,pacheco2024langevin,weron2017ergodicity,munoz2023quantitative}. In the same manner, the correlation coefficient between components can fluctuate}, that is, $\rho$ may not be constant. In the financial dataset analyzed in the previous section, we limited ourselves to a 2-year period to avoid such a situation. However, data for both indices have been available for more than 67 years. Thus, in a general situation, one might find that the correlation coefficient depends on time, where $\rho(t)$ describes the local correlation between coordinates. We can express the correlation through the time-dependent cross-covariance function
\begin{align} 
\langle \dd B_1(t) \dd B_2(t) \rangle = 2\sqrt{D_1 D_2}\rho(t) \dd t. 
\end{align}
Naturally, in a more general situation, $D_1$ and $D_2$ may also depend on time, but we restrict ourselves to the case of constant diffusivity in this work.

\begin{figure}[ht!]
\centering
\includegraphics[width=0.9\linewidth]{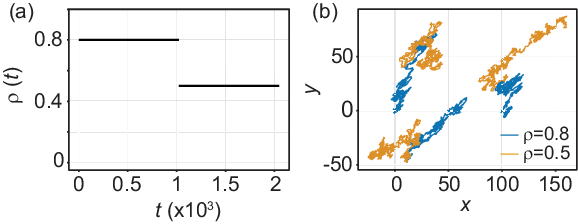}
 \caption{Numerical simulations of two-dimensional BM with a correlation that exhibits a change point. (a) $\rho(t)$ is a stepwise constant with a change point at $t_0=1024$. (b) Three sample trajectories. The blue and yellow colors correspond to the higher and lower correlations, respectively. The considered length of the simulated trajectories is $2048$. For clarity of presentation, two trajectories were shifted by 50 and 100 units to the right.}
     \label{fig:8}
\end{figure}

In order to showcase a case where the correlation changes, we consider a simple situation where the correlation is a step function
\begin{align*}
    \rho(t) = \begin{cases}
        \rho_1, \quad \textrm{ for } t \leq t_0,\\
        \rho_2, \quad \textrm{ for } t > t_0,\\
    \end{cases}
\end{align*}
where $\rho_1, \rho_2$ are constants within $(-1, 1)$ and there is a single change point that occurs at time $t_0 > 0$.
%
Figure \ref{fig:8} illustrates this case with $\rho_1=0.8, \rho_2 = 0.5$ and a change point at $t_0=1024$. Figure~\ref{fig:8}(a) shows $\rho$ as a function of time with levels $\rho_1 = 0.8$ and $\rho_2 = 0.5$. Figure~\ref{fig:8}(b) presents three sample trajectories in which the time domains with $\rho_1 = 0.8$ and $\rho_2 = 0.5$ are colored blue and yellow, respectively. For clarity of presentation, two trajectories were shifted by 50 and 100 units to the right. The underlying randomness of the corresponding trajectories is the same as in Figure \ref{fig:trajs1}.

\subsection{Detecting non-constant correlation}

\begin{figure}[ht!]
\centering
\includegraphics[width=0.9\linewidth]{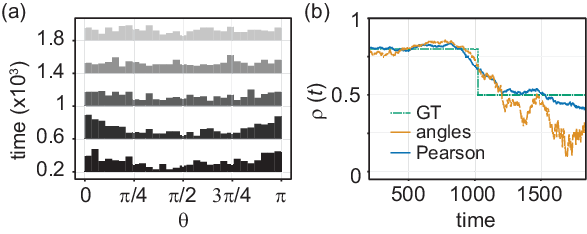}
 \caption{Analysis of numerical simulations with stepwise constant correlation coefficient. (a) Local distribution of the turning angles at different times. The darker the histogram's bars, the earlier the considered data window. (b) Estimated correlation coefficient $\rho$ from the Pearson correlation and the least-squares fit of the distribution of turning angles as a function of time based on eq. (\ref{eq:angle_mod_final}). The dash-dotted line shows the ground truth (GT). In both cases, the used sliding window has a length of $400$.}
     \label{fig:9}
\end{figure}

An important question is how to detect fluctuations in the correlation $\rho$ from experimental data. A way to achieve this goal involves the analysis of the turning angle distribution and how it changes over time. In cases where only a single trajectory is available, e.g., financial datasets, we can infer the correlation coefficient ${\rho}$ or the distribution of turning angles by employing a sliding window of length $w$.
To present this approach, we consider simulated data with $\rho(t)$ corresponding to the stepwise constant function presented in Figure \ref{fig:8}(a). In our numerical simulations, we use a moving window of length $w=400$, while the length of the sample trajectory is $2^{11}$. 
The results obtained are presented in Figure \ref{fig:9}, where panel (a) shows the evolution of the histogram of the turning angles over time. The darker colors of the histograms correspond to earlier parts of the trajectory, while time $t$ on the vertical axis corresponds to the index of the midpoint of the sliding window, i.e. the histogram at $t=600$ corresponds to the turning angles calculated at times $t=401, 402, \ldots, 800$.
Figure \ref{fig:9}(b) shows the correlation $\widehat{\rho}(t)$ estimated in two different ways. First, the standard Pearson estimator of the correlation coefficient $\rho$ is obtained as a function of time $t$ corresponding to the data points in the window $\{t-w/2+1, \ldots, t+w/2\}$, in a similar way as in Figure~\ref{fig:9}(a). Second, $\widehat{\rho}(t)$ is calculated from the distribution of turning angles within the same window, using a least-squares method (denoted in Figure~\ref{fig:9} as "angles"). Here, the distribution of the turning angle is fitted to the function presented in eq.~ (\ref{eq:angle_mod_final}). Both estimation methods give a similar result. {For $\rho=0.8$, the estimation of correlation coefficient from the turning angle distribution performs very well, yielding accurate results. However, as the correlation decreases ($\rho=0.5$) the estimation error increases significantly. In such cases, using a larger time window is necessary to achieve more accurate correlation estimates. A larger window would be needed for accurate correlation estimation. In any case, it is clear from the data that the distribution of turning angles, using a small time window, undergoes a noticeable change around $t=1000$, accompanied by a decrease in correlation.}

\subsection{Example: Financial data}
\begin{figure}[ht!]
    \centering
    \includegraphics[width=0.9\linewidth]{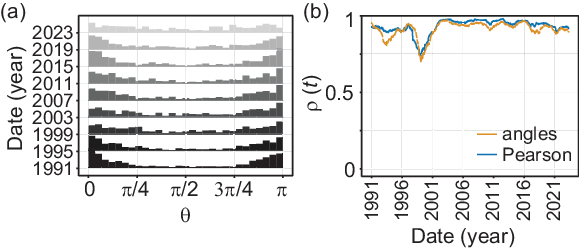}
    \caption{Analysis of extended financial dataset. (a) Distribution of turning angles evaluated at different times. The darker the histograms, the older the considered data window. (b) Estimated correlation coefficient $\rho$ from the Pearson correlation and the least-squares fit of the distribution of turning angles as a function of time based on eq. (\ref{eq:angle_mod_final}). In both cases, the sliding window used to obtain the correlation has a length of $500$ days.}
    \label{fig:10}
\end{figure}

We analyze the correlation and the distribution of the turning angle for the financial dataset (DJIA and S\&P500) with an extended time period ranging from January 1, 1990, to May 1, 2024. Following the approach used for the numerical simulations (Figure \ref{fig:9}), we performed a comparable analysis for long financial data, and the results are presented in Figure \ref{fig:10}. We considered daily recordings and a window length $w=500$ days, approximately 2 years of data. Figure \ref{fig:10}(a) presents histograms of the turning angles of windows in different years. We see that, especially in 1994 and 1998, the distributions exhibit marked fluctuations, indicating varying $\rho$ in time.
Figure \ref{fig:10}(b) presents the comparison of the two approaches to estimate $\rho(t)$. Both the classical Pearson estimator and the least-squares fit of the distribution of turning angles yield similar results. It is worthwhile to note the significant drop around the year 1998. This decrease in correlation between the two indices corresponds to the year when DJIA recorded a 2\textsuperscript{nd}-worst daily loss, resulting from Wall Street being overwhelmed by the 1998 Russian financial crisis. 

\section{Conclusions}

In this article, we have introduced and analyzed a model for two-dimensional Brownian motion with spatial correlations, characterized by the correlation coefficient $\rho$ and the diffusivities in each direction $D_1$ and $D_2$. We derived some of the key properties, such as the autocovariance and cross-covariance functions, the mean square displacement, and the turning angle distribution given in a general case by a convolution. We demonstrated that while the MSD exhibits the expected linear behavior typical of normal diffusion, spatial correlation introduces significant changes in the distribution of turning angles and the overall trajectory patterns. The distribution of turning angles offers valuable insights into the correlation between trajectory components. While such correlations can also be analyzed {in terms of} the Pearson correlation coefficient of the increments projected onto different {axes}, turning angles serve as a crucial measure in the study of dynamics within complex systems. {In particular, while Pearson's coefficient is sensitive to an arbitrary change in orientation of the coordinate system, the turning angle statistics are not. Thus, the turning angles probe the internal dynamics of the physical system, independent of the orientation of the coordinate system of the laboratory
frame. These statistics are also} useful for evaluating temporal correlations, which may arise in contexts such as viscoelastic environments or fractal geometries. Consequently, it is important to understand how spatial correlations (as studied in this work) influence the distribution of turning angles, as well as their impact on other metrics like the mean squared displacement (MSD) and covariance functions, and how it changes over time in heterogeneous processes.

Through numerical simulations, we illustrate the impact of correlation coefficients on sample trajectories, PDF contour plots, and displacement histograms. Numerical simulations confirmed that spatial correlations markedly affect the statistical properties of motion, providing a deeper insight into the behavior of the introduced model.
Historically, turning angles have been underutilized, leading to a gap in the comprehensive understanding of the data. {The turning angle distribution is slowly gaining more attention and has been employed to study a diverse array of systems including the compartmentalization and clustering in the plasma membrane \cite{sadegh2017plasma, mosqueira2020antibody}, the motion of telomeres in the nucleus \cite{hidalgo2024directed}, the dynamics of insulin granules in the cytoplasm and contractile cytoskeletal networks on model membranes \cite{burov2013distribution}, turbulent flows \cite{bos2015angular}, and even the trajectories of football players \cite{kadoch2017directional}.} By incorporating turning angles alongside correlations, we gain a better understanding of the data and improve our understanding of directional information.
To provide an additional focus on the usefulness of the model, we applied it to real-world datasets, namely financial indices (Dow Jones Industrial Average index and Standard and Poor's 500) and physical systems (polystyrene beads in aqueous solution). The results highlighted the model's applicability across fields, reinforcing its utility in describing complex systems with spatial correlations.

Furthermore, we recognized that assuming a constant correlation coefficient can be misleading in real-world data. By introducing a time-dependent correlation coefficient $\rho(t)$, we captured the dynamic nature of the correlations. We showed that local correlations could vary significantly over time through both simulated and financial data, affecting the behavior of the system under study. This variability in the correlation coefficient was evident from the changes in the turning angle distributions and the corresponding estimates of $\rho(t)$. Using a sliding window approach allowed us to detect and quantify these variations, providing a more accurate representation of the underlying stochastic processes. Recognizing and accounting for time-varying correlations not only enhances the theoretical understanding of Brownian motion with spatial coefficients but also offers practical insights in fields where such models are applied.  
Overall, our study provides a comprehensive framework for understanding multidimensional Brownian motion with spatial correlations, offering valuable tools for both theoretical analysis and practical applications in various scientific and financial fields.

Turning angles offer a compelling alternative to relying solely on correlation and variance measurements for several key reasons. They present an intriguing characteristic, providing insight into directional changes within data sets. Their interpretation is the most natural for the physical type of data describing positions, but even in other fields, it finds useful applications, e.g., in the presented financial data. Using the turning angles adds depth to our understanding of the underlying patterns, especially in contexts with crucial temporal or spatial dependencies. 

Future work could incorporate a memory structure for each of the coordinates, e.g. by using a two-dimensional fractional Brownian motion model \cite{lavancier2009covariance, amblard2011identification,MarajZygmat,krapf2019spectral}. The analytical results for the turning angles would be more complicated to obtain, as such a model incorporates temporal correlations. However, the extension to higher dimensions of correlated BM is straightforward. The turning angles are still random variables with PDF as given in eq. (\ref{eq:angles_general}), although dependent on a larger number of parameters.
Other possible extensions relate to processes with an infinite second moment, such as L\'evy flights \cite{chechkin2006fundamentals,palyulin2019first} and fractional L\'evy stable motion \cite{Burneckietal2010,Janczuraetal2022}, and to processes with non-stationary increments, such as multifractional Brownian motions \cite{ayache2004identification,wang2023,imfbm,balcerek2023},   {heterogeneous diffusion processes with transitions points \cite{montiel2006quantitative,akimoto2017detection,pacheco2024fractional,pacheco2024langevin}}, and continuous time random walks \cite{weietal11,kutner2017continuous}. Similarly, other dependence measures, such as covariation or codifference, play a similar role to the correlation coefficient in this framework \cite{st}.  

\section{Acknowledgments}
This work was supported by the National Science Foundation Grant 2102832 (to DK) and the National Science Centre, Poland, projects 2020/37/B/HS4/00120 (to AW) and 2023/07/X/ST1/01139 (to MB).

\bibliography{bibliography}

\end{document}